# DOWN CONVERTER DEVICE COMBINING RARE-EARTH DOPED THIN LAYER AND PHOTONIC CRYSTAL FOR C-SI BASED SOLAR CELL

Thierry Deschamps[1,2], Emmanuel Drouard[1,2], Romain Peretti[1,2], Loic Lalouat[1,2], Erwann Fourmond[1,3], Alain Fave[1,3], Antoine Guille[4], Antonio Pereira[4], Bernard Moine[4], Christian Seassal[1,2,3]
[1]Institut des Nanotechnologies de Lyon (INL), UMR 5270, CNRS-INSA-ECL-UCBL, France
[2]Ecole Centrale de Lyon, 36 Avenue Guy de Collongue, 69134 Ecully Cedex, France
[3]INSA de Lyon, Bat. Blaise Pascal, 7 Avenue Capelle, 69621 Villeurbanne, France
[4]Institut Lumière Matière, UMR 5306 Université Lyon 1-CNRS, Université de Lyon 69622 Villeurbanne cedex, France

ABSTRACT: The aim of the study is to develop ultra-compact structures enabling an efficient conversion of single high energy photon (UV) to two lower energy photons (IR). The proposed structure combines rare-earths doped thin layer allowing the down-conversion process with a photonic crystal (PhC), in order to control and enhance the down-conversion using optical resonances. On the top of the rare-earths doped layer, a silicon nitride (SiN) 2D planar PhC is synthesized. For that, SiN is first deposited by PECVD. After holographic lithography and reactive ion etching, a periodic square lattice of holes is generated on the SiN layer. The PhC topographical parameters as well as the layers thickness are optimized using Finite-Difference-Time-Domain simulations. The design and realization of such PhC-assisted down-converter structures is presented. Optical simulations demonstrate that the PhC leads to the establishment of resonant modes located in the underneath doped layer, allowing a drastic enhancement of the absorption of the rare-earth ions without disturbing the transmission in the visible and near-IR parts of the spectrum, hence demonstrating the relevance of such an approach.

## 1 INTRODUCTION

With the aim of reaching the greed parity, many routes are currently explored to simultaneously enhance the PV cells yield while reducing the use of raw material, meaning the PV cell cost. Concerning the crystalline silicon-based solar cells, which represent the main part of the marketed cells, different solutions are proposed to decrease optical losses: anti-reflect coating, nanostructured thin c-Si layer [1,2], frequency converter devices [3]… This last concept, which should allow to overcome the Shockley-Queisser limit [4], can be divided into three parts: up-conversion, down-shifting and down-conversion [5-7].

The down-conversion process, by which one high energy photon is converted into two lower energy photons, is extremely promising. Indeed, such a process allows avoiding carriers thermalization, a strong limitation of c-Si PV cell in the UV-blue range, using a rare-earths doped thin layer located on top of the PV cell. However, due to the low absorption of the rare-earths doped thin layer, the external quantum yield of the down-conversion needs to be increased.

In this paper, we propose an optically resonant structure, combining a rare-earth doped thin layer with a 2D planar PhC, with the aim to realize an efficient down-converter device for c-Si PV cells. The concept of this original device is developed section 2. Results of optical simulations showing the potentiality of our approach (section 3) and the nano-fabrication steps leading to the complete elaboration of the structure (section 4) are described.

## 2 THE DOWN-CONVERTER DEVICE

### 2.1 General concept

The goal of the down-converter proposed is to efficiently convert, with an internal quantum yield higher than one, UV-blue incident photons to the near-IR range where c-Si PV cells are highly sensitive. To reach this goal, the UV absorption of the rare-earth ions in the active thin layer has to be increased strongly. We proposed that this active layer be held as an integral part of a UV resonant structure, thanks to nanophotonics architecture. In this paper, only the enhancement of the UV absorption is addressed.

### 2.2 Choice of the materials

Concerning the active doped layer, we have chosen the couple of rare-earth ions {$Pr^{3+}$;$Yb^{3+}$} in a $CaYAlO_4$ matrix to induce the down-conversion. However, because the absorption cross-section of $Pr^{3+}$ is extremely low, we add a third ion, $Ce^{3+}$, which has a much higher (but still low) absorption cross-section than $Pr^{3+}$ thanks to its 4f-5d parity-allowed transition [8]. In $CaYAlO_4$, $Ce^{3+}$ absorbs UV light between 320 and 420 nm, and can efficiently transferred its energy to $Pr^{3+}$ by non-radiative transfers. The goal of the PhC is then to enhance, by the introduction of slow Bloch modes, the lifetimes of the incident photons in the structure in the absorption domain of $Ce^{3+}$. We note that the $Yb^{3+}$ emission rate at 980 nm is expected to be linear with the absorption rate of $Ce^{3+}$.

Different constraints guide the choice of the material constituting the PhC. This material has to be transparent as far as possible in the whole solar spectral range to avoid useless absorption, must have a high refractive index to improve light trapping in the structure, and must be compatible with usual etching processes. All these requirements are fulfilled by silicon nitride material which has been chosen to constitute the planar PhC, located on the top of the doped active thin layer.

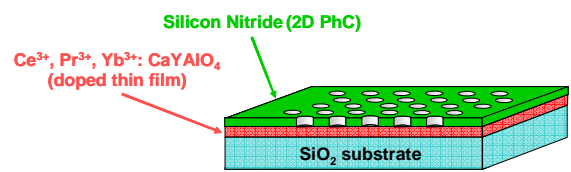

**Figure 1:** Schematic view of the down-converter device proposed

## 3  OPTICAL SIMULATIONS

The optical simulations have been carried out using a 3D Finite-Differential Time-Domain (FDTD) approach. The implemented optical properties (n,k dispersion) of each material have been experimentally determined by means of spectroscopic ellipsometry measurements. The source used is a normal incident plane wave in the range 320-420 nm for the calculation of $Ce^{3+}$ absorption, and 400-1150 nm for the calculation of the global transmission of the structure.

The thickness of the $CaYAlO_4$ doped layer has been set as low as 100 nm. The geometrical parameters of the SiN PhC have been optimized in order to reach the greater $Ce^{3+}$ absorption while retaining an excellent transmission of the whole structure.

The optimised parameters obtained using a periodic square lattice of air holes are a lattice parameter of 224 nm, an air filling factor of 0.40, and a SiN thickness of 90 nm.

Results concerning the absorption enhancement in the UV range and the global transmission of this specific structure are presented section 3.1 and 3.2 respectively.

### 3.1 UV absorption

In order to determine the $Ce^{3+}$ ions absorption in the 100 nm $CaYAlO_4$ thin film, we have calculated the transmission spectra at the SiN/CYA and CYA/$SiO_2$ interfaces. The simulated absorption spectra of $Ce^{3+}$, with and without patterning the SiN layer, are displayed Fig.1. We note that the fraction absorbed by the host $CaYAlO_4$ matrix herself has been removed.

The SiN patterning as a PhC introduce two resonant modes in the range 320-420 nm, which are mainly localized in the doped layer. These two modes lead to a drastic enhancement of the $Ce^{3+}$ absorption.

The $Ce^{3+}$ ions integrated absorption with the PhC is 10 times higher than the $Ce^{3+}$ absorption without SiN patterning.

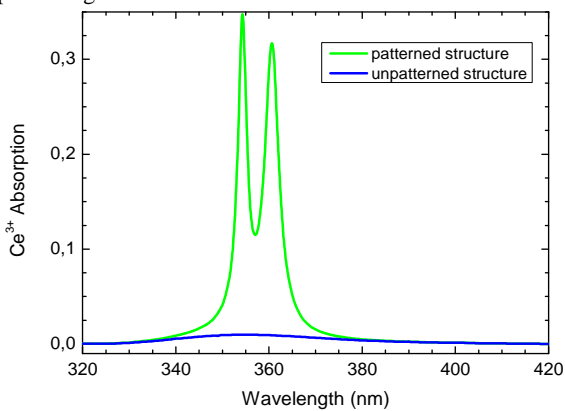

**Figure 2:** $Ce^{3+}$ absorption in the $CaYAlO_4$ layer with the unpatterned (blue line) and the patterned (green line) SiN layer.

### 3.2 Transmission of the structure

The transmission spectra for the patterned and unpatterned structures have been calculated between 400 nm and 1150 nm, i.e in the useful spectral range of a c-Si PV cell. The solar spectral photon flux (AM 1.5) have been then calculated through these two stacks and compared to the transmission through a silica glass slide for a first test (Fig.3).

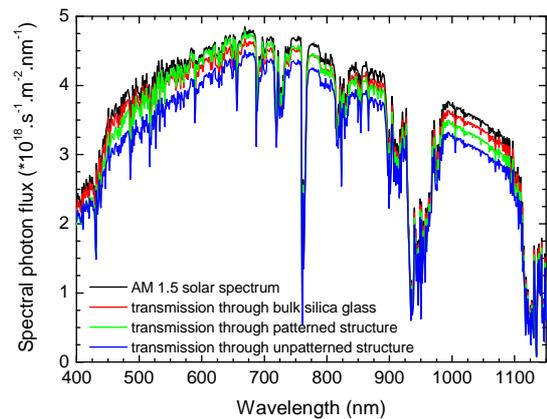

**Figure 3:** AM 1.5 solar photon flux spectrum (black line) and its transmission through a silica glass slide (red line), through the patterned structure proposed (green line), and through the unpatterned structure (blue line)

The transmission results demonstrate that the PhC behaves as an excellent anti-reflector in the whole range of the solar spectrum, especially between 550 nm and 850 nm where the solar photon flux is maximal. The integrated photon flux through the patterned structure is comparable to the one through a silica glass slide. On the other hand, the unpatterned structure strongly degrades the transmission.

## 4  ELABORATION OF THE STRUCTURE

The fabrication of the whole stack can be divided into two main parts: thin films deposition (section 4.1) and patterning of the silicon nitride layer as a PhC (section 4.2). The SEM characterization of the samples is presented section 4.3.

### 4.1 Thin films deposition

Starting from pure silica glass substrates, the deposition of the rare-earth doped thin layers has been realised using a conventional Pulse Laser Deposition (PLD) technique. For that, a pulsed UV laser beam (KrF at $\lambda$=193 nm) was focalised on a $CaYAlO_4$ target doped with Ce, Pr and Yb (atomic concentration 0.5%, 1.5% and 10% respectively) in $O_2$ atmosphere (P=$10^{-2}$ mBar). The samples were then heated in a furnace at 800°C during 6 hours under nitrogen atmosphere to avoid rare-earth oxydation. This thermal treatment shifts the thin doped layer from an amorphous to a micro-crystalline state and increases the absorption efficiency of the rare-earth ions.

On the top of this layer has been then deposited the silicon nitride layer using low-frequency Plasma-Enhanced Chemical Vapor Deposition (PECVD) at 370°C. The stoichiometry was controlled adjusting the ammonia-to-silane gas flow ratio $NH_3$/$SiH_4$.

### 4.2 Silicon nitride layer patterning

In order to generate a square lattice of air holes in the SiN layer, holographic lithography process with a 10 mW UV laser operating at 266 nm was used. The samples were first coated with a NEB 22 negative tone chemically amplified resist, and then insolated during 140 s in two orthogonal directions (2D PhC). The angle between the laser beam and the normal of the sample surface was adjusted to get the desired parameter lattice. After a 120 s post-exposure bake at 92°C, the resist was developed using a MF702 solution. The pattern was then transferred from the resist to the SiN layer using Reactive Ion Etching with a gas mixture of $SF_6$ (15sccm) and $CHF_3$ (40sccm). The remaining resist was removed by an $O_2$ plasma.

4.3 SEM characterization

To demonstrate the feasibility of the SiN 2D PhC elaboration, SEM images showing the SiN patterned layer directly deposited on a silica glass substrate are displayed Fig.4 (on these samples, the $CaYAlO_4$ doped layer has not been deposited). Fig.4a shows the periodic square lattice of air holes patterning a SiN layer of 100 nm thickness obtained after the process steps described previously. The air filling factor which is very sensitive to the insolation and post-exposure bake duration, has been measured at 0.38, i.e closed to the desired value of 0.40. To easily check the quality of the etching process, a 1D PhC sample has been elaborated and then cut along an axis orthogonal to the patterning. A side view of this 1D PhC is presented Fig.4.b. Those SEM images demonstrate the high quality of the etching process, namely concerning the sidewall verticality.

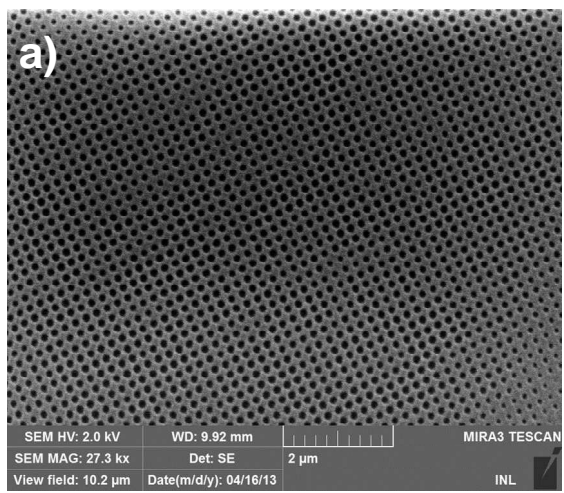

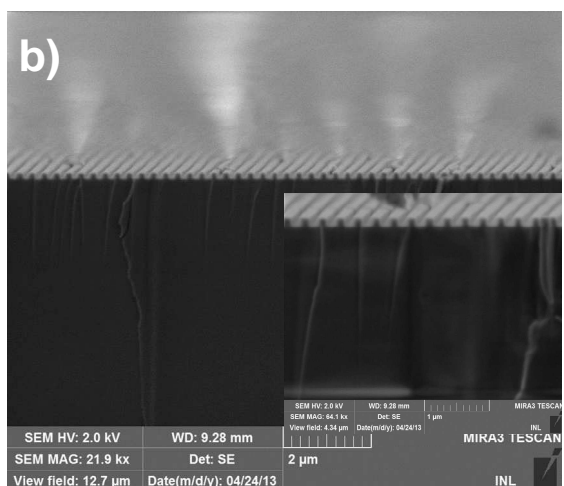

**Figure 4:** a) Top SEM view of the SiN 2D PhC after holographic lithography and dry RIE etching process
b) Side views of the SiN 1D PhC showing the high quality of the etching process.

5   CONCLUSION

We have presented a new concept, combining rare-earth doped thin film and PhC, in order to increase the external efficiency of the down-conversion for PV applications. As a result of series of FDTD simulations, we have demonstrated that the proposed design allows enhancing the integrated absorption of $Ce^{3+}$ ions in the UV-blue range by about a factor of 10. This clearly shows the high potentialities of the proposed approach. Moreover, optical simulations have demonstrated that the SiN PhC also behaves as an excellent anti-reflector in the whole range of the solar spectrum. In further works, we will address not only the absorption control in the UV domain, but also the emission dynamics and radiation pattern in the near-IR.

The technical feasibility of the SiN 2D PhC with specific parameters has also been demonstrated. The experimental validation of the down-conversion enhancement, by means of photoluminescence measurements, is currently in progress.

ACKNOWLEDGMENT

The authors are grateful to the Program "Investissements d'Avenir"- launched by the French Government and operated by the National Research Agency (ANR) - for financial support to the LabEx iMUST of Université de Lyon. The technology platform NanoLyon is also gratefully acknowledged for its technical support.